\documentclass[twocolumn,preprintnumbers,amsmath,amssymb]{revtex4}
\usepackage{dcolumn}
\usepackage{bm}

\begin{document}

\title{Why graphene conductivity is constant: scaling theory consideration}
\author{V. Vyurkov$^{{\rm 1}}$}
\email{vyurkov@ftian.ru}
\author{V. Ryzhii$^{{\rm 2}}$}%
\email{v-ryzhii@u-aizu.ac.jp}
\affiliation{%
$^{{\rm 1}}$Institute of Physics and Technology RAS, Moscow 117218, Russia\\
$^{{\rm 2}}$University of Aizu, Aizu-Wakamatsu 965-8580, Japan,
and Japan Science and Technology Agency, CREST, Tokyo 107-0075,
Japan }
\date{March 25, 2008}

\begin{abstract}
In the recent paper [arXiv: cond-mat/0802.2216, 15 Feb 2008],
Kashuba argued that the intrinsic conductivity of graphene
independent of temperature originated in strong electron-hole
scattering. We propose a much more explicit derivation based on a
scaling theory approach. We also give an explanation of a rapid
increase in graphene conductivity caused by applied gate voltage.
\end{abstract}
\maketitle


Experimental investigations of graphene have revealed a quite
amazing peculiarity of graphene resistivity. There is almost no
dependence of a maximum resistivity of graphene $\rho _{{\rm
m}{\rm a}{\rm x}}$ on temperature (between $0.3K$ and $300K$)
\cite{geim2007}. This is really intriguing because at the same
time the carrier density varies by six orders of magnitude.

The maximum resistivity (or intrinsic resistivity) arises at zero
gate voltage in a gated graphene structure. The resistivity
approximately equals $\rho _{{\rm m}{\rm a}{\rm x}} \approx 6
kOhm$. There appears a natural temptation to bind this value to
two conductance quanta as graphene has a twofold degeneracy of its
band structure \cite{geim2007}. The conductance quantum for
spin-unpolarized electron gas equals $G _{{\rm 0}}=2e^{{\rm 2}}/h
=(12kOhm)^{{\rm -} {\rm 1}}$. The obtained $6 kOhm$ roughly equals
$h/4e^{{\rm 2}}$, that is, right two conductance quanta. Here we
suggest that the coincidence is purely accidental and, actually,
$\rho _{{\rm m}{\rm a}{\rm x}}$ is determined by electron-hole
scattering and bound to the velocity $v_{{\rm F}}$ characterizing
the graphene band-structure and effective permittivity $\kappa
_{{\rm e}{\rm f}{\rm f}}$.

The maximum resistivity (intrinsic resistivity) corresponds to an
utterly ambipolar graphene plasma when the electron density
$\Sigma _{{\rm e}}$ and the hole density $\Sigma _{{\rm h}}$
coincide: $\Sigma _{{\rm e}}=\Sigma _{{\rm h}}$. In the situation
the electron-hole scattering is the strongest one.

In the recent paper \cite{kashuba}, Kashuba also asserted that the
intrinsic conductivity of graphene independent of temperature
originated in strong electron-hole scattering. We propose below a
much simpler derivation based on scaling theory approach.

The graphene conductivity $\sigma $ caused by Coulomb scattering
among electrons and holes could be plainly supplied by the
expression

\begin{equation}
\label{eq1} \sigma = {\frac{{\kappa _{{\rm e}{\rm f}{\rm f}}^{2}}}
{{e^{2}}}}f,
\end{equation}

\noindent where $\kappa _{{\rm e}{\rm f}{\rm f}}$ is an effective
permittivity and $f$ is a so far unknown function of electron-hole
system in graphene. In the equation (\ref{eq1}) a proper
dependence upon the elementary charge $e$ and the effective
permittivity $\kappa _{{\rm e}{\rm f}{\rm f}}$ for Born
approximation of Coulomb scattering is factorized. As for
screening in graphene layer, it was recently theoretically
investigated in \cite{shytov, fogler, biswas}. This screening
turned out to be determined by a weak function of a dimensionless
parameter

\begin{equation}
\label{eq2} \alpha = {\frac{{e^{2}}}{{\kappa _{{\rm e}{\rm f}{\rm
f}} \hbar v_{F}}} }.
\end{equation}

Here we omit this kind of screening. Really, screening by gates
might be substantial were they placed fairly close to a graphene
sheet.

The factor $f$ in equation (\ref{eq1}) can solely depend on
intrinsic graphene parameters and the Plank constant $h$. In the
case when electron and hole densities coincide $\Sigma _{{\rm
e}}=\Sigma _{{\rm h}}=\Sigma _{{\rm T}}$ the graphene plasma is
fully characterized by an only parameter besides the Fermi
velocity $v_{{\rm F}}$. It could be a mean momentum $p_{{\rm T}}$,
or a mean wave vector $k_{T} = {\frac{{p_{T}}} {{\hbar}} }$, or
temperature $T \approx {\frac{{\hbar v_{F} k_{T}}} {{k}}}$ (where
$k$ is the Boltzmann constant), or density $\Sigma _{T} \approx
k_{T}^{2} $. All those parameters are directly bound to each
other. Therefore, the only possible combination looks like

\begin{equation}
\label{eq3} \sigma = {\frac{{\kappa _{{\rm e}{\rm f}{\rm f}}^{2}
hv_{F}^{2}}} {{e^{2}}}} = {\frac{{2\kappa _{{\rm e}{\rm f}{\rm
f}}^{2} v_{F}^{2}}} {{G_{0}}} }.
\end{equation}

Here we neglect a numerical factor of the order of unity which
could be derived via thorough consideration of electron-hole
scattering \cite{kashuba} or determined from experimental data
\cite{geim2007}.

The only temperature-dependent parameter $k_{{\rm T}}$ describing
graphene plasma can not be included. Therefore, graphene
conductance becomes independent of temperature. Moreover, it is
readily seen that the graphene conductivity turned out to be
inversely proportional to the conductance quantum $G_{{\rm
0}}=2e^{{\rm 2}}/h =(13kOhm)^{{\rm -} {\rm 1}}$, in spite of
former expectations \cite{geim2007}.

It is expedient to estimate the value of $\sigma $ for typical
graphene structures. When a graphene sheet is sandwiched between
two dielectrics with permittivity $\kappa _{{\rm 1}} $ and $\kappa
_{{\rm 2}}$, the effective permittivity is

\begin{equation}
\label{eq4} \kappa _{{\rm e}{\rm f}{\rm f}} = {\frac{{\kappa _{1}
+ \kappa _{2}}} {{2}}}.
\end{equation}

Therefore, in our opinion, the crucial experiment to verify the
above proposed theory lies in changing the permittivity of
adjacent dielectrics. In most of so far made experiments a
graphene layer was deposited on SiO$_{{\rm 2}}$ surface. The
effective permittivity in the case is equal to $\kappa_{{\rm
e}{\rm f}{\rm f}} \approx 2.5$. The velocity $v_{{\rm F}}=10^{{\rm
8}}cm/s$ in CGS system of units corresponds to the conductance
equal to $(3kOhm)^{{\rm -} {\rm 1}}$. Substituting all that in
equation (\ref{eq3}) one obtains $\sigma \approx (4kOhm)^{{\rm -}
{\rm 1}}$. It is a very good agreement with the experimental value
equal to $(6kOhm)^{{\rm -} {\rm 1}} $\cite{geim2007} allowing for
a numerical coefficient in relation (\ref{eq3}) was ignored.

All said above belongs to an intrinsic graphene conductivity when
both electron and hole densities are exactly equal to each other:
$\Sigma _{{\rm e}}=\Sigma _{{\rm h}}=\Sigma _{{\rm T}}$. It seems
interesting to apprehend what occurs beyond this equality which
can be broken by a gate voltage. At fairly high gate voltage when
$\Sigma _{{\rm e}}\gg \Sigma _{{\rm h}}$ or vice versa, the
graphene plasma is almost unipolar and electron-hole scattering is
no more important. In that case, the conductivity is likely
governed by scattering on charged defects \cite{geim2007,hwang} as
there is no temperature dependence of mobility evidenced by
experiments. Obviously, for perfect graphene structure phonon
scattering dominates and governs graphene conductivity
\cite{vasko}. Probably, a carrier mobility can achieve values
above $200~000~ cm^{{\rm 2}}/ V s$ instead of nowadays $5~000 -
15~ 000~ cm^{{\rm 2}}/ V s$ at room temperature \cite{morozov}.

The explanation of constant value of intrinsic graphene
conductivity put forward above implies that the electron-hole
scattering predominates over all other mechanisms of scattering
even at very low temperature ($1-10K$) when electron and hole
density is very small ($10^{{\rm 9}}-10^{{\rm 1}{\rm 0}}cm^{{\rm
-} {\rm 2}})$. At first glance, this definitely contradicts
another intriguing feature of graphene conductivity observed in
experiments \cite{geim2007}, namely, a rapid increase in graphene
conductivity with rising gate voltage. The defect concentration
consistent with measured values of mobility in the range of $5~000
-15~000 ~cm^{{\rm 2}}/ V s$ was estimated as $10^{{\rm 1}{\rm
2}}~cm^{{\rm -} {\rm 2}}$ \cite{geim2007}. To cope with this
contradiction we suggest that charged defect concentration depends
on position of the Fermi level in graphene, or the Fermi energy
$\varepsilon _{F}$. For instance, such defects may arise at the
graphene/SiO$_{{\rm 2}}$ interface. If number of defects per unit
energy and area is constant and equals $n_{i}$ their space
concentration is

\begin{equation}
\label{eq5} \Sigma _{i} = n_{i} \varepsilon _{F} \approx n_{i}
v_{F} \hbar \sqrt {\Sigma } ,
\end{equation}

\noindent where $\Sigma $ is a majority carrier density (electrons
or holes). For intrinsic graphene state the Fermi energy in the
relation (\ref{eq5}) should be replaced by thermal energy $kT$.
This supposition substantiates that charged defect density can be
smaller than that of carriers even at low temperatures.

In conclusion, we assert that maximum (intrinsic) resistivity of
graphene originates in a strong electron-hole scattering when both
electron and hole densities coincide. It does not depend on
temperature and can be manipulated by permittivity of surrounding
dielectrics and screening of metallic gates.

\bigskip
\begin{acknowledgments}
The authors thank F. T. Vasko for valuable information.
\end{acknowledgments}

\bibliography{graphene_conductivity_brief}

\begin{thebibliography}{8}
\expandafter\ifx\csname natexlab\endcsname\relax\def\natexlab#1{#1}\fi
\expandafter\ifx\csname bibnamefont\endcsname\relax
  \def\bibnamefont#1{#1}\fi
\expandafter\ifx\csname bibfnamefont\endcsname\relax
  \def\bibfnamefont#1{#1}\fi
\expandafter\ifx\csname citenamefont\endcsname\relax
  \def\citenamefont#1{#1}\fi
\expandafter\ifx\csname url\endcsname\relax
  \def\url#1{\texttt{#1}}\fi
\expandafter\ifx\csname urlprefix\endcsname\relax\def\urlprefix{URL }\fi
\providecommand{\bibinfo}[2]{#2}
\providecommand{\eprint}[2][]{\url{#2}}

\bibitem[{\citenamefont{Geim and Novoselov}(2007)}]{geim2007}
\bibinfo{author}{\bibfnamefont{A.~K.} \bibnamefont{Geim}} \bibnamefont{and}
  \bibinfo{author}{\bibfnamefont{K.~S.} \bibnamefont{Novoselov}},
  \bibinfo{journal}{Nature Materials} \textbf{\bibinfo{volume}{6}},
  \bibinfo{pages}{183} (\bibinfo{year}{2007}).

\bibitem[{\citenamefont{Kashuba}(15 Feb 2008)}]{kashuba}
\bibinfo{author}{\bibfnamefont{A.}~\bibnamefont{Kashuba}},
  \emph{\bibinfo{title}{Conductivity of the defectless graphene}}
  (\bibinfo{year}{15 Feb 2008}), \eprint{cond-mat/0802.2216}.

\bibitem[{\citenamefont{Shytov et~al.}(23 Jul 2007)\citenamefont{Shytov,
  Katsenelson, and Levitov}}]{shytov}
\bibinfo{author}{\bibfnamefont{A.~V.} \bibnamefont{Shytov}},
  \bibinfo{author}{\bibfnamefont{M.~I.} \bibnamefont{Katsenelson}},
  \bibnamefont{and} \bibinfo{author}{\bibfnamefont{L.~S.}
  \bibnamefont{Levitov}}, \emph{\bibinfo{title}{Vacuum polarization and
  screening of supercritical impurities in graphene}} (\bibinfo{year}{23 Jul
  2007}), \eprint{cond-mat/0705.4663v2}.

\bibitem[{\citenamefont{Fogler et~al.}(29 Aug 2007)\citenamefont{Fogler,
  Novikov, and Shklovskii}}]{fogler}
\bibinfo{author}{\bibfnamefont{M.~M.} \bibnamefont{Fogler}},
  \bibinfo{author}{\bibfnamefont{D.~S.} \bibnamefont{Novikov}},
  \bibnamefont{and} \bibinfo{author}{\bibfnamefont{B.~I.}
  \bibnamefont{Shklovskii}}, \emph{\bibinfo{title}{Screening of hypercritical
  charge in graphene}} (\bibinfo{year}{29 Aug 2007}),
  \eprint{cond-mat/0707.1023v3}.

\bibitem[{\citenamefont{Biswas and Suchdev}(3 Sept 2007)}]{biswas}
\bibinfo{author}{\bibfnamefont{R.~R.} \bibnamefont{Biswas}} \bibnamefont{and}
  \bibinfo{author}{\bibfnamefont{S.}~\bibnamefont{Suchdev}},
  \emph{\bibinfo{title}{Coulumb impurity in graphene}} (\bibinfo{year}{3 Sept
  2007}), \eprint{cond-mat/0706.3907v2}.

\bibitem[{\citenamefont{Hwang et~al.}(2007)\citenamefont{Hwang, Adam, and
  Sarma}}]{hwang}
\bibinfo{author}{\bibfnamefont{E.~H.} \bibnamefont{Hwang}},
  \bibinfo{author}{\bibfnamefont{S.}~\bibnamefont{Adam}}, \bibnamefont{and}
  \bibinfo{author}{\bibfnamefont{S.~D.} \bibnamefont{Sarma}},
  \bibinfo{journal}{Phys. Rev. Lett.} \textbf{\bibinfo{volume}{98}},
  \bibinfo{pages}{186806} (\bibinfo{year}{2007}).

\bibitem[{\citenamefont{Vasko and Ryzhii}(22 Aug 2007)}]{vasko}
\bibinfo{author}{\bibfnamefont{F.~T.} \bibnamefont{Vasko}} \bibnamefont{and}
  \bibinfo{author}{\bibfnamefont{V.}~\bibnamefont{Ryzhii}},
  \emph{\bibinfo{title}{Voltage and temperature dependence of conductivity in
  gated graphene}} (\bibinfo{year}{22 Aug 2007}),
  \eprint{cond-mat/0708.2976v1}.

\bibitem[{\citenamefont{Morozov et~al.}(2008)\citenamefont{Morozov, Novoselov,
  Katsnelson et~al.}}]{morozov}
\bibinfo{author}{\bibfnamefont{S.~V.} \bibnamefont{Morozov}},
  \bibinfo{author}{\bibfnamefont{K.~S.} \bibnamefont{Novoselov}},
  \bibinfo{author}{\bibfnamefont{M.~I.} \bibnamefont{Katsnelson}},
  \bibnamefont{et~al.}, \bibinfo{journal}{Phys. Rev. Lett.}
  \textbf{\bibinfo{volume}{100}}, \bibinfo{pages}{016602}
  (\bibinfo{year}{2008}).

\end{thebibliography}

\end{document}